\documentclass{article}
\def\DH{\rm I\kern-1.5pt\rm H\kern-1.5pt\rm I}

\input epsf

\def\DR{\rm I\kern-1.45pt\rm R}
\def\DC{\kern2pt {\hbox{\sqi I}}\kern-4.2pt\rm C}

\newcommand{\ba}{\begin{array}}
\newcommand{\ea}{\end{array}}
\newcommand{\be}{\begin{equation}}
\newcommand{\ee}{\end{equation}}
\newcommand{\bea}{\begin{eqnarray}}
\newcommand{\eea}{\end{eqnarray}}
\newcommand{\bi}{\begin{itemize}}
\newcommand{\ei}{\end{itemize}}

\def\theequation{\arabic{section}.\arabic{equation}}
\begin{document}
\thispagestyle{empty}
\begin{center}
{\bf \Large  Second Hopf map and supersymmetric mechanics with Yang monopole}\\
\vspace{0.5 cm} {\large M. Gonzales$^1$, Z. Kuznetsova$^{1,2}$, A.
Nersessian$^{3,4}$, F. Toppan$^{1}$ and V. Yeghikyan$^{4}$}
\end{center}

$\;^1${\it CBPF, Rua Dr. Xavier Sigaud 150, cep 22290-180, Rio de Janeiro (RJ), Brazil}

$\;^2${\it UFABC, Rua Catequese 242, Bairro Jardim,  cep 09090-400, Santo Andr\'e (SP), Brazil}

$\;^3${\it Artsakh State University, 5 Mkhitar Gosh St. ,
  Stepanakert, Armenia}

$\;^4${\it Yerevan State University, 1 Alex Manoogian St.,   Yerevan, 
Armenia}
\begin{abstract}
We propose to use the second Hopf map for the reduction (via $SU(2)$ group action)
of the eight-dimensional ${\cal N}=8$ supersymmetric mechanics
 to five-dimensional supersymmetric systems specified by the
 presence of an $SU(2)$ Yang monopole. For our purpose we develop the relevant reduction procedure. The reduced system is characterized by its invariance under the
 ${\cal N}=5$ or ${\cal N}=4$ supersymmetry generators
 (with or without an additional conserved BRST charge operator) which commute with the $su(2)$ generators.
\end{abstract}
\section{Introduction}
Recently, in a serie of papers, new non-linear one-dimensional
supermultiplets have been suggested
\cite{{Ivanov:2003nk},{IKL},{k1}}. They were used to construct new
models of two- and three-dimensional  ${\cal N}=4$ supersymmetric
mechanics. An important peculiarity of these models is the
appearance  of external magnetic fields preserving the
supersymmetry of the system \cite{IKL},\cite{bbkno}-\cite{bko}.
Those, contain, as particular cases, important systems like the
${\cal N}=4$ supersymmetric Landau model \cite{IKL} and  the
${\cal N}=4$ supersymmetric
 multi-center MICZ-Kepler systems, both conventional \cite{kno} and spherical
 \cite{bko}\footnote{ The MICZ-Kepler system is the generalization of the Kepler system specified
 by the presence of a Dirac monopole and inherits  the hidden symmetry of the Kepler system.
 It was invented independently by Zwanziger
 and by McIntosh and Cisneros in Refs. \cite{Z, mic}.}.
 Some unexpected phenomenon has been observed: it was
found that in the two-dimensional case the nonlinear (chiral) supermultiplet provides a wide freedom in the construction of
supersymmetric extensions of given bosonic systems, parameterized by
an arbitrary holomorphic function (``$\lambda(z)$-freedom")\cite{freedom}.\par
It was shown in \cite{Gates} and \cite{pashtop} that all linear one-dimensional
 ${\cal N}=4$ multiplets are related and can be derived
from the so-called ${\cal N}=4$  ``root multiplet" or ``minimal length multiplet" (i.e. the multiplet
possessing no bosonic auxiliary degrees of
freedom).
An important step in understanding the nature of nonlinear multiplets has been done in \cite{root}. The
nonlinear chiral multiplet used in the construction of
two-dimensional supersymmetric mechanics possesses the  $({2,4,2})$
components content\footnote{ We follow the nowadays standard
convention in the literature of denoting with
  $({k,n, n-k})$  the supermultiplets with ${k}$ physical bosons, ${n}$ physical fermions and ${ n-k}$ auxiliary bosons.},
 while the three-dimensional systems are built with a
multiplet possessing $({3,4,1})$ components content.
 The minimal length multiplet from which nonlinear multiplets are obtained possesses a
 $({4,4,0})$ fields content.
Looking at the construction of \cite{root} one can observe that it
is  is related with the reductions associated with the first Hopf
map ${S}^3/{S}^1={S}^2$ and with, respectively, the
Kustaanheimo-Stiefel transformation \cite{ks}. The relation of the
mentioned  procedures with the first Hopf map becomes especially
transparent after their reformulation in the Hamiltonian language
\cite{bny}. It is therefore not surprising that the reduced
three-dimensional system is specified by the presence of a Dirac
monopole field, while the two-dimensional one is specified by the
presence of a constant electric field. We further notice that the
performed reductions do not change the number of fermionic degrees
of freedom, i.e.  they are straightforward extensions
 of the purely bosonic reduction procedures to supersymmetric systems.

Different supersymmetric extensions (for various values of ${\cal N}$) admit unique minimal length
linear multiplets with a
given number of bosonic and fermionic degrees of freedom. The relevant cases here are, for
${\cal N}=2$, the  $({2,2,0})$ root supermultiplet, for ${\cal N}=8$ the $({8,8,0})$ supermultiplet \cite{pashtop}.
There is no doubt
that the first supermultiplet can be related with the zero-th Hopf map ${S}^1/{S}^0={S}^1$, while
the latter is related with the
second Hopf map ${S}^7/{S}^3={S}^4$.
Since ${S}^0={\bf Z}_2$, the reduction associated with the zero-th Hopf map does not
change
the number of physical degrees of freedom;
at the classical level it corresponds to a plain coordinate transformation even if,
at the quantum-mechanical level, it yields the presence of
magnetic fluxes generating spin $1/2$ \cite{ntt}. Looking at the number of components of
the $({8,8,0})$  multiplet,
one could naively expect the existing $({4,8,4})$ and $({5,8, 3})$  ${\cal N}=8$ nonlinear
multiplets being obtained
from $({8,8,0})$ via a second Hopf map reduction.
It is likely that a second Hopf map reduction applied to the
system with $({8,8,0})$ multiplet (the Hamiltonian reduction is
assumed via the action of the ${S}^3=SU(2)$ group), would produce
a five-dimensional supersymmetric mechanics model with Yang
monopole and (upon a further fixation  of the radius) a
four-dimensional supersymmetric mechanics system with BPST
instanton. Indeed, when involving only the bosonic part of the
system, the  $SU(2)$ reduction produces a five-dimensional model
in the presence of a Yang monopole; in \cite{iwai}, such reduction
was used for constructing the five-dimensional MICZ-Kepler system
($SU(2)$-Kepler system) from an eight-dimensional system.\par
 The construction of  ${\cal N}>4$ supersymmetric extensions of the systems with
 Yang monopole is clearly an important  task.
 As mentioned before, systems of this type are important not only from a purely field-theoretical context,
  but also in applications to condensed matter, e.g. in the theory of the four-dimensional Hall
effect (which is  formulated on the ground of a four-dimensional Landau problem, namely a
 particle on a four dimensional sphere moving in the presence of a BPST instanton field
 generated by the Yang monopole located at the center of the sphere) \cite{4h}.
  Therefore, with the supersymmetric  four-dimensional Landau problem at hand, one can develop the theory
  of the four-dimensional quantum Hall effect, in the spirit of \cite{hasebe}.

On the other hand, the systems produced by existing $({4,8,4})$ and $({5,8, 3})$  ${\cal N}=8$
    linear multiplets do not contain any external gauge field.
 However, the extension of the reduction procedure of
the $({4},{4},{0})$ multiplet to the $({ 8},{ 8},{0})$ (which
supposes the transition from the first Hopf map to the second one)
and the construction of the associated nonlinear supermultiplets,
is not a trivial task. In contrast with the  reduction  of
$({4},{4},{0})$ by the $U(1)$ group action,
 the $({ 8},{8},{0})$ multiplet must be reduced by the non-Abelian $SU(2)$ group
 action.
 Such a reduction implies the ``elimination" of the three external bosonic degrees of freedom only in a limiting case
(when the values of $SU(2)$ generators are equal to zero). In a general position part of the initial degrees
of freedom results in internal  degrees of freedom of the isospin particle interacting with a Yang monopole.
In the ``supermultiplet language" this means that the auxiliary fields of the resulting nonlinear
supermultiplet should contain some ``emergent dynamics"; indeed, they are not ``auxiliary" in a strong sense.
Some other points need to be clarified: performing the reduction of the $({4},{4},{0})$ multiplet
to the nonlinear ones, the authors of \cite{root} added to the initial system, by hands, a Fayet-Iliopoulos extra-term.
It has the two aims of
providing the final system with a nonlinearity property and with the presence of an external magnetic field.
Naively, it
would seem that the relation of the mentioned supermultiplets is not so straightforward.
From the above construction it
 is not clear which sort of Fayet-Iliopoulos term should be added to the system with $({8},{8},{0})$
 multiplet for producing a
 lower-dimensional system with Yang monopole.
 Finally, one can suppose, from group-theoretical considerations, that it would not
 be possible to reduce  all initial ${\cal N}=8$ supersymmetries to  low dimensions.

The goal of the present paper is to clarify the listed questions and, consequently, develop the necessary tools
for the reduction of the  ${\cal N}=8$ supersymmetric mechanics with  $({8},{ 8},{ 0})$
to five (four)-dimensional mechanics in
presence of Yang monopoles (BPST instantons) which possess the extended supersymmetry\footnote{
To our knowledge no  ${\cal N}> 4$ supersymmetric
mechanical model with a non-Abelian gauge  field has been realized.
In a context different from ours we mention
the recent paper \cite{ped} where the authors derived the $SU(2)$ supersymmetric Yang-Mills quantum mechanics from dimensional reduction of $d=3,4,6,10$ superYang-Mills theories and related the Berry holonomy with the Hopf maps.\par One of the main differences with our approach
is the fact that we are investigating the most general supersymmetric quantum mechanics obtained from the minimal, irreducible supermultiplets.
}.

For this purpose we formulate at first the reduction procedures
associated with the first and second Hopf maps.
 We show that there is no need to add the Fayet-Iliopoulos-like term to the initial system:
 the full time-derivative term arises
naturally within a consistent reduction procedure. Also, we propose a geometric
construction of the transmutation
of the ``seemingly auxiliary" degrees of freedom in isospin degrees of freedom. Let us
mention that we formulate
 the reduction associated with the second Hopf map by using the quaternionic language.
  The simpler case related with the first Hopf map can be easily recovered by the obvious
 replacement of the  quaternionic quantities with complex numbers.
An algebraic understanding of the nature of the Hopf maps leaves to no surprise that important differences
are encountered between the first and the second Hopf map. We consider the consequences of these reductions for
supersymmetric mechanics.

The first Hopf map induces, starting from an ${\cal N}=4$ supersymmetric quantum mechanics with $4$ (target) dimensions,
${\cal N}=4$ s´persymmetric quantum mechanical systems with either $2$ or $3$ target dimensions.
The reason lies in the fact that the initial ${\cal N}=4$ superalgebra commutes with the generator of the $S^1=U(1)$ symmetry
(the defining bundle in the first Hopf map), by whose action
the reduction is performed.
In the second Hopf map one must reduce the ${\cal N}=8$ supersymmetric mechanics
constructed with the $({8,8,0})$ supermultiplet in terms of the action of the $SU(2)=S^3$ group
(the defining bundle in the second Hopf map). These generators do not commute with the whole
set of ${\cal N}=8$ supersymmetry algebra, but at most with its ${\cal N}=5$ subalgebra.
The reduced system,
in the presence of a Yang monopole, is fully characterized by its invariance under the ${\cal N}=5$  $SU(2)$-invariant supersymmetry generators.
It is even possible, under some condition on the initial eight-dimensional system,
to combine the fifth supersymmetry generator with a conserved pseudosupersymmetry operator and produce a reduced
${\cal N}=4$ supersymmetric quantum mechanical model and an additional odd nilpotent (BRST-type) symmetry.
We  restrict ourselves to the presentation of the general procedure and the listed statements, postponing a detailed
analysis for forthcoming publications.

The paper is arranged as follows.

In the Second Section we present an explicit description of the first and second Hopf maps in terms  needed for our purposes.

In the Third section  we employ the Hopf maps
to reduce the four-/eight-dimensional bosonic  systems to lower dimensional systems with magnetic/SU(2) monopoles.

In the Fourth Section we apply these reduction procedures to the supersymmetric mechanics constructed
in terms of, respectively, the $({4,4,0})$ and $({8,8,0})$ minimal length supermultiplets and discuss the
associated resulting supermultiplets of the reduced systems.

\setcounter{equation}{0}
\section{Hopf maps}

The Hopf maps (or Hopf fibrations) are the fibrations of the
sphere over a sphere, $ {  S}^{2p-1}/{  S}^{p-1}={  S}^{p}$,
$p=1,2,4,8$. These fibrations reflect the existence of real
($p=1$), complex ($p=2$), quaternionic ($p=4$) and octonionic
($p=8$) numbers.

We are interested in the so-called  first and second Hopf maps:
\be
 {  S}^3/{  S}^1={  S}^2\quad ({\rm first}\;{\rm Hopf}\;{\rm map}), \qquad
{  S}^7/{  S}^3={  S}^4\quad ({\rm second}\;{\rm Hopf}\;{\rm map}).
  \label{hm0}
 \ee
Let us describe them   in explicit terms. For this purpose, we
consider the  functions ${\bf x}(u_\alpha,{\bar u}_\alpha),
x_{p+1}(u_\alpha,{\bar u}_\alpha)$

\be {\bf x}=2{\bf \bar u}_1{\bf u}_2 ,\quad x_{p+1}={\bf \bar
u}_1{\bf u}_1-{\bf \bar u}_2{\bf u}_2, \label{hm}\ee where ${\bf
u}_1,{\bf u}_2$ are complex numbers for $p=2$ case (first Hopf map)
and quaternionic numbers for the $p=4$ case (second Hopf map). One
can consider them as coordinates of the $2p$-dimensional  space
$\DR^{2p}$ ($p=2$ for ${\bf u}_{1,2}$ complex numbers; $p=4$ for
${\bf u}_{1,2}$ quaternionic numbers). In all  cases  $x_{p+1}$ is a
real number while ${\bf x}$ is, respectively, a complex number
($p=2$) or a quaternionic one ($p=4$), \be {\bf x}\equiv x_{p}
+\sum_{k=1,\ldots, p-1}{\bf e}_k x_k,\qquad \ee where ${\bf e}_k =
{\bf i}$, ${\bf i}^2=-1$ for $p=2$, and ${\bf e}_k= ({\bf i},{\bf
j},{\bf k})$, ${\bf e}_i{\bf e}_j=-\delta_{ij}+\varepsilon_{ijk}
{\bf e}_k$ for $p=4$.

 Hence, $(x_{p+1}, {\bf x})$ parameterize the $(p+1)$-dimensional space $\DR^{p+1}$.

 The functions  ${\bf x}, x_{p+1}$ remain invariant under the
 transformations
\be {\bf u}_\alpha\to {\bf{G}} {\bf u}_\alpha ,\quad {\rm
where}\quad {\bar {\bf {G}}}{\bf{G}}=1 \Rightarrow \left\{
\begin{array}{lll}
{\bf {G}}=\lambda_1+{\bf i}\lambda_2 &
|\lambda_1|^2+|\lambda_2|^2=1 &{\rm for}\quad p=2 \cr {\bf
G}=\lambda_1+{\bf i}\lambda_2+{\bf j}\lambda_3+{\bf k}\lambda_4 &
|\lambda_1|^2+\ldots+|\lambda_4|^2=1&{\rm for}\quad p=4 .
\end{array}\right.
\ee Therefore,  ${\bf G}$ parameterizes the spheres $S^{p-1}$ of
unit radius. Taking into account the  isomorphism
 between these spheres and the groups, ${  S}^1=U(1)$, ${  S}^3=SU(2)$,
we get that (\ref{hm})  is invariant under $G-$group
transformations (where $G=U(1)$ for $p=2$, and $G=SU(2)$ for
$p=4$), and that it defines the fibrations \be \DR^4/{
S}^1=\DR^3,\quad\DR^8/{  S}^3=\DR^5. \ee
 One could immediately
check that the following equation holds: \be r^2\equiv {\bf \bar
x}{\bf x}+x^2_{p+1}=( {\bf\bar u}_{1}{\bf u}_1+{\bf \bar u}_2{\bf
u}_2)^2\equiv R^4. \label{hm1}\ee
 Thus, defining the $(2p-1)$-dimensional sphere in
$\DR^{2p}$ of radius $R$, ${\bf\bar u}_\alpha{ \bf u}_\alpha =R^2$,
we will get the $p$-dimensional sphere in $\DR^{p+1}$ with radius
$r=R^2$, i.e. we obtain the Hopf maps (\ref{hm0}).

The expressions (\ref{hm}) can be easily inverted by the use of \be
{\bf u}_\alpha={\bf g} r_\alpha ,\qquad {\rm where }
\quad r_1=\sqrt{\frac{r+x_{p+1}}{2}},\quad  r_2\equiv
r_+=\frac{{\bf x}}{\sqrt{2(r+x_{p+1})}},\quad
,\quad {\bar {\bf g}}{\bf g}=1 .
\label{inve}\ee
It follows from the last equation in (\ref{inve}) that ${\bf g}$
parameterizes the $(p-1)$-dimensional sphere of unit radius.
Let us give the description of first and second Hopf maps in internal terms, using the
decomposition
 $\DR^{2p}= \DR^1\times {  S}^{2p-1}$,  $\DR^{p+1}=\DR^1\times {  S}^p$, and paramerizing ${  S}^p$
 by inhomogeneous projective coordinates
 \be
 z= \frac{{\bf \bar u}_1 {\bf u}_2}{{\bf\bar u}_1{\bf u}_1}, \quad \Rightarrow \quad |{\bf u}_1|^2 =\frac{r}{1+{\bar z}z}.
 \label{5}\ee
Hence, we get \be {\bf u}_1=\frac{{\bf g}{\sqrt r} }{\sqrt{1+{\bar
z}z}},\qquad {\bf u}_2={\bf u}_1z=\frac{{\bf g}\sqrt{r} z
}{\sqrt{1+{\bar z}z}}
 \label{51}\ee
For $r={\rm const}$ we get the description of ${  S}^{2p-1}$ in
terms of the coordinates of the base manifold ${  S}^{p} $ and of
the fiber coordinates ${\bf g}$. The internal coordinate $z$ of
the sphere ${  S}^p$ is related with the Cartesian coordinates of
the ambient space $\DR^{p+1}$ (\ref{hm}) as follows \be {\bf
x}={r}{\bf h}_+,\quad x_{p+1}={r}{ h}_{p+1},\qquad {\bf
h}_+=\frac{2z}{1+\bar z z}, \quad { h}_{p+1}=\frac{1-\bar z
z}{1+\bar z z}. \label{kil}\ee

For ${  S}^1$ the group element and
the corresponding left-invariant one-form can be presented as
follows \be {  S}^1 \;:\quad {\bf g}= {\rm  e}^{{\bf i} \varphi},
\quad {\overline {\bf g}}{d {\bf g}}= {\bf i} d\varphi ,\quad
\varphi\in [0, 2\pi ) \label{s1}\ee Hence,  the ambient
coordinates of the ${  S}^3$ sphere of unit radius are related
with the internal coordinates of ${  S}^1$ and ${  S}^2$ by
(\ref{51}), where  we put $r=1$ and ${\bf g}={\rm e}^{{\bf
i}\varphi}$.\par

 In quaternionic case we get  the following
expressions for the $SU(2)$ group element and its left-invariant
form \be {  S}^3\;:\quad {\bf g}={\rm e}^{{\bf i}
\gamma}\frac{1+{\bf j} z}{\sqrt{1+z\bar z}}, \quad {\overline {\bf
g}}{d {\bf g}}=\Lambda_3{\bf i} +\Lambda_+ {\bf j},\qquad
\Lambda_+=(\Lambda_2+{\bf i}\Lambda_1), \label{lam1}\ee where \be
\Lambda_3=h_{3}d\gamma +\frac{{\bf i}}{2}\frac{\bar{z} d z-z
d\bar{z}}{1+z\bar{z}} \qquad \Lambda_+={\bf i}{\bf h}_+ d\gamma
+\frac{d \bar z}{1+z\bar{z}}
 \qquad i,j,k=1,2,3.\label{13}\ee
Here $h_3, {\bf h_\pm}$ are  the Euclidean coordinates of the
ambient space $\DR^{3}$ given by (\ref{kil}): simultaneously they
play the role of Killing potentials of the K\"ahler structure on
$S^2$.

The vector fields dual to the above one-forms  look as follows
\be\label{vec1} {\bf V}_3= \frac{\partial}{\partial\gamma}+2{\bf
i} \left(z\frac{\partial}{\partial z} -
\bar{z}\frac{\partial}{\partial \bar{z}}\right) ,\qquad {\bf
V}_+=\frac{\partial}{\partial \bar z}
+{z}^2\frac{\partial}{\partial z} -{\bf
i}\frac{z}{2}\frac{\partial}{\partial\gamma} ,\qquad {\bf V}_-=
\overline{{\bf V}}_+: \label{vec}\ee \be \Lambda_3({\bf
V}_3)=\Lambda_\pm({\bf V}_\pm )=1,\qquad \Lambda_{\pm}({\bf V}_\mp
)=\Lambda_\pm ({\bf V}_3)= \Lambda_3({\bf V}_\pm)=0. \ee
Let us also write down the following expressions\be-\left(
{\bar{\bf g}}{d {\bf g}}\right)^2=\Lambda_i\Lambda_i=
\left(d\gamma -\frac{{\bf i}}{2}\frac{\bar{z}
dz-zd\bar{z}}{1+z\bar{z}}\right)^2+ \frac{dzd\bar z}{(1+z\bar
z)^2}.\label{qar} \ee

We also need another $SU(2)$ group element parameterizing the
sphere ${  S}^3$ and ``commuting" with (\ref{lam1}):
 \be {\bf \widetilde g}=\frac{1+{\bf j} z}{\sqrt{1+z\bar{z}}}{\rm e}^{-{\bf i}
\gamma}, \qquad  {\bf \bar g}{\bf \bar{\widetilde g}}{\bf  g}{\bf
\widetilde g}=1. \label{gti}\ee The corresponding left-invariant
forms are given by the expressions \be {\bar {\bf\widetilde g}}{d
{\bf\widetilde g}}= {\widetilde\Lambda}_3 {\bf i}
+{\widetilde\Lambda}_+ {\bf j},\qquad
{\widetilde\Lambda}_+={\widetilde\Lambda}_2+{\bf
i}{\widetilde\Lambda}_1,\qquad {\widetilde\Lambda}_3= d\gamma
+\frac{{\bf i}}{2}\frac{zd\bar z-\bar z dz}{1+z\bar z} \qquad
{\widetilde\Lambda}_+= \frac{{\rm e}^{2{\bf i} \gamma}d\bar
z}{1+z\bar z},\label{13_1}
\label{lam2}\ee while the vector fields dual to these forms look
as follows:\be\label{vec2}
 {\bf
U}_3= -\frac{\partial}{\partial\gamma}, \qquad {\bf{U}}_+= {\rm
e}^{-2 {\bf i}\gamma}\left(\left(1+z \bar
z\right)\frac{\partial}{\partial \bar z}+\frac{{\bf i} z}{2}
\frac{\partial}{\partial \gamma}\right), \qquad {\bf{U}}_-=
\overline{{\bf{U}}}_+: \ee \be {\widetilde\Lambda}_3({\bf
U}_3)={\widetilde\Lambda}_\pm({{\bf U}}_\pm )=1,\qquad
{\widetilde\Lambda}_{\pm}({{\bf U}}_\mp )={\widetilde\Lambda}_\pm
({{\bf U}}_3)= {\widetilde\Lambda}_3({{\bf U}}_\pm)=0. \ee
From the second expression in (\ref{lam2}) follows the
commutativity of the ${\bf V}_a$ and ${\bf U}_a$ fields. This pair
forms the  the $so(4)=so(3)\times so(3)$ algebra of isometries of
the ${  S}^3$ sphere.
 \be
[{\bf V}_i,{\bf V}_j]=2\varepsilon_{ijk}{\bf V}_k,\qquad [{\bf
U}_i,{\bf U}_j]=2\varepsilon_{ijk}{\bf U}_k,\qquad [{\bf V}_i,{\bf
U}_j]=0,\qquad\qquad i,j,k=1,2,3. \label{comm} \ee The commutativity
of ${\bf V}_i$ and ${\bf U}_i$ plays a key role in our further
considerations. Notice also that we can pass from the
parametrization (\ref{lam2}) to (\ref{lam1}) via the $ z\rightarrow
\tilde{z} {\rm e}^{-2 {\bf i}\tilde{\gamma}}$, $
\gamma=-\tilde{\gamma}$ transformation.

For our further considerations this is all we need to know from
the Hopf maps.

 \setcounter{equation}{0}
\section{Reduction}

Let us consider a free particle on the $2p$-dimensional space
equipped with the $G$-invariant conformal flat metric. Taking into
account the expressions (\ref{inve}) we can represent its
Lagrangian as follows
$$
{\cal L}_{2p}=g({\bf \overline u}\cdot {\bf u}){\bf\dot{\overline
u}}_\alpha {\bf\dot{u}}_\alpha=
$$
\be
 =g(r_\pm,r_1)\left({\dot r}_+{\dot r}_- +{\dot r}^2_1
+{\dot r}_-\overline{{\bf g}}{\dot {\bf g}}{ r}_+ -r_-
\overline{{\bf g}}{\dot {\bf g}}{\dot r}_+ - r(\overline{{\bf g}}{\dot
{\bf g}})^2\right)= g\left({\dot r}_+{\dot r}_- +{\dot
r}^2_1\right)- gr{\Lambda}_{i}A_{i} +gr{ \Lambda}_{i}{ \Lambda}_{i}
,
 \label{lag0}\ee
 Here and in the following
${\Lambda}_i$ are  defined by (\ref{s1}) for $p=2$, and by
(\ref{13}) for $p=4$, with the differentials replaced by the time
derivatives, while
\be A_i\equiv \frac{{\dot r}_- {\bf e}_ir_+ -{r}_- {\bf e}_i{\dot
r}_+}{r}= \frac{{\bf\dot{{\overline x}}}{\bf e}_i{{{\bf
x}}}-{{\bf{\overline  x}}} {\bf e}_i{\bf\dot{ x}}
}{2r(r+{x}_{p+1})}. \label{monop}\ee
We have used the identity  $r_+r_- +r^2_1=r$ and
the notations $r_-={\overline r_+ }$,  ${\bf{\overline u}}\cdot
{\bf u}\equiv {\bf {\overline u}}_\alpha \cdot {\bf u}_\beta $.

One can see, for the $p=2$ case (the complex numbers) that ${\cal
A}$ defines a Dirac monopole potential
\be {A}_i= A_D=\frac{x_1{\dot x}_2-x_2{\dot
x}_1}{r(r+x_3)}. \ee
In the $p=4$ case (the quaternionic numbers) $A_i$ defines the
potential of the the $SU(2)$ Yang monopole.
 The explicit formulae for
$A_i$ in terms of the real coordinates $x_1,\ldots , x_5$ (where ${\bf x}=x_4+{\bf e}_i x_i$, $x_5$) look as follows:
$$
A_i=\frac{\eta^i_{ab}x_a\dot{x}_b}{r\left(r+x_5\right)},
\qquad\eta^i_{ab}=\delta_{ia}\delta_{4b}-\delta_{4a}\delta_{ib}-\varepsilon_{iab 4},
$$
where $\eta^i_{ab}$ is the t'Hooft symbol, and $a,b=1,2,3,4$.

The Lagrangian  (\ref{lag0}) is manifestly invariant under the
$G-$group action.

In the  $p=2$ case the generator of the $G=U(1)$ group is given by
the vector field ${\bf V}=\partial/\partial\varphi$:
 indeed, taking into account (\ref{s1}), one can see that, for $p=2$, $\varphi$ is a cyclic variable in (\ref{lag0}).

In the $p=4$ case  the generators of the $G=SU(2)$  group are
given by the vector fields $ {\bf U}_i $ (\ref{vec2}).

By making use of the Noether constants of motion we can decrease
the dimensionality of the system.

In the $p=2$ case we have a single   Noether constant of motion
defined  by the vector field dual to the left-invariant form
${\Lambda}={\dot\varphi}$; this is precisely the momentum
conjugated to $\varphi$, which appears in the Lagrangian
(\ref{lag0}) as a cyclic variable.
 Hence, excluding  this variable, we shall get, for  $p=2$, a three-dimensional system.

On the other hand, in the $p=4$ case, thanks to the non-Abelian
nature of the $G=SU(2)$ group, only the $\gamma$ variable is a
cyclic one, even if $z,{\bar z}$  appear in the Lagrangian
(\ref{lag0}) without time-derivatives too. It is therefore
expected that in this second case the reduction procedure would be
more complicated. In contrast with the Hamiltonian reduction
procedure, the Lagrangian reduction is a less common, or at least
a less developed, procedure which deserves being done with care.

For this reason, we shall describe the Lagrangian counterparts of
the Hamiltonian reduction procedures separately for both the $p=2$
and the $p=4$ cases.

\subsection{The $U(1)$ reduction}

Let us consider  the reduction of the four-dimensional particle
given by the Lagrangian (\ref{lag0}) to a three-dimensional
system. Taking into account the expression (\ref{s1}), we can
re-write the Lagrangian as follows:
\be {\cal L}=g\left({\dot
r}_+{\dot r}_- +{\dot r}^2_1 -r\dot\varphi {\cal A}_D +
r{\dot\varphi}^2\right). \label{l4}\ee
Since $\varphi$ is a cyclic
variable, its conjugated momentum is a conserved quantity
 \be
p_\varphi=\frac{\partial {\cal L}}{\partial\dot\varphi}=-rg {\cal
A}_D+ 2gr\dot\varphi\quad\Leftrightarrow\quad
\dot\varphi=\frac 12\left(\frac{p_\varphi}{gr}+ {\cal A}_D\right). \ee
Naively one could
expect that the reduction would require fixing the value of the
Noether constant and substituting corresponding expression for
$\dot\varphi$  in the Lagrangian (\ref{l4}). However, acting in
this way, we shall get a three-dimensional Lagrangian without a
linear term in the velocities, i.e. without a magnetic field (of
the Dirac monopole).
This would be in obvious contradiction with
the result of the Hamiltonian reduction of the four-dimensional
system via the $U(1)$ group action.
The correct
reduction procedure looks as
follows.
 At first  we have to  replace the Lagrangian (\ref{l4}) by the following,
variationally equivalent, one (obtained by performing the Legendre
transformation for $\dot\varphi$):
 \be \widetilde{\cal
L}=p_\varphi {\dot\varphi}-\frac{p_\varphi}{2}{\cal A}_D
-\frac{p^2_\varphi}{4rg}-\frac{g r}{4} A_D^2+ g\left({\dot r}_+{\dot r}_-
+{\dot r}^2_1\right). \label{l5}\ee Indeed, varying the
independent variable $p_\varphi$, we shall arrive to the initial
Lagrangian.

The isometry of the Lagrangian (\ref{l5}), corresponding to the
$U(1)$-generator
 ${\bf V}=\frac{\partial}{\partial\varphi}$, is given by the same vector field. It
defines the Noether constant of motion $p_\varphi$.

 {
Upon fixing the value of the Noether constant
\be
p_\varphi=2s,
\ee the
first term of the new Lagrangian transforms as a full time
derivative and can therefore be ignored.}

As a result, we shall get the following three-dimensional
Lagrangian \be {\cal L}_3=g\left({\dot r}_+{\dot r}_- +{\dot
r}^2_1\right)-s{\cal A}_D-\frac{g r}{4}{\cal A}_D^2 -\frac{s^2}{rg}=
\frac{{\widetilde g}{\dot x}_\mu {\dot x}_\mu}{2} -s{\cal A}_D
-\frac{s^2}{2r^2{\widetilde g}},\qquad {\widetilde g}\equiv \frac{g}{2r}.\qquad \mu=1,2,3. \label{lred}\ee Clearly, it
describes the motion of a particle moving in a three-dimensional
space equipped by the metric ${\widetilde
g}_{\mu\nu}=\frac{g}{2r}\delta_{\mu\nu}$ in presence  of a Dirac
monopole generating a magnetic field with strength \be {\vec
B}=\frac{s{\vec x}}{{\widetilde g}x^3}. \ee

Let us notice the appearance, in the reduced system, of the specific centrifugal term
$s^2/2r^2{\widetilde g}$. For spherically symmetric systems this term provides a minor modification of the solutions
of the initial system (without monopole) after incorporating the Dirac monopole: at the classical level it yields only the rotation
of the orbital plane to the $\arccos s/J$ angle \cite{lnp} and, at the quantum level, the shift of the validity range of the
orbital momentum $J$  from $[0,\infty)$  $[|s|,\infty )$ \cite{mny}.
Schwinger \cite{schw} incorporated  by hands, for the first time such a term in planar systems (${\widetilde g}=1$)
 with Dirac monopole.

The above construction
corresponds to the bosonic part of the reduction of the
four-dimensional ${\cal N}=4$ supersymmetric mechanics to a
three-dimensional ${\cal N}=4$ supersymmetric mechanics considered
in \cite{root}.
A further reduction of the system to two dimensions corresponds to
a system with a nonlinear chiral multiplet $({ 2},{ 4},{
2})$, obtained by fixing the ``radius" $r=const$. Since the Dirac
monopole potential ${\cal A}_D $ does not depend on $r$, we shall
get a two-dimensional system moving in the same  magnetic
field. It applies in particular to the particle on the sphere
moving in a constant magnetic field (the Dirac monopole is located
at the center of the sphere), i.e. the Landau problem on sphere.

Let us also mention the serie of papers \cite{gauging}, where the  $U(1)$ reduction procedure
of the supersymmetric Lagrangian mechanics has been performed by the use of a specific ``gauging" procedure, which seemingly
 could be reduced, in the bosonic sector,  to the above presented one.
\subsection{The $SU(2)$ reduction}

In the case of the second Hopf map we have to reduce the
Lagrangian (\ref{lag0}) with $p=4$ via the action of the $SU(2)$
group expressed by the vector fields (\ref{vec2}). Due to the
non-Abelian nature of the $SU(2)$ group the system will be reduced
to a five (or higher)-dimensional one.

For a correct reduction procedure we have to replace the initial
Lagrangian by one which is variationally equivalent, extending the
initial configuration space with new variables, $\pi$, $\bar\pi$,
$p_\gamma$, playing the role of conjugate momenta to $z$, $\bar
z$, $\gamma$. In other words, we will  replace the sphere ${ S}^3$
(parameterized by $z$, $\bar z$, $\gamma$) by its cotangent bundle
$T^*{  S}^3$ parameterized by the coordinate $z$, $\bar z$,
$\gamma$, $\pi$, $\bar\pi$, $p_\gamma$.
 Let us further define, on $T^*{  S}^3$,
the Poisson brackets given by the relations \be \{\pi,
z\}=1,\qquad\{\bar\pi, \bar z\}=1,\qquad\{p_\gamma, \gamma\}=1.
\label{pb}\ee We introduce the  Hamiltonian generators $P_a$
corresponding to  the vector fields (\ref{vec1}) (replacing  the
derivatives entering the vector fields ${\bf V}_a$  by the half of
corresponding momenta) \be
 P_+=\frac{P_2-{\bf i}
P_1}{2}=\frac{\pi+\bar{z}^2 \bar\pi}{2}-{{\bf i} \bar{z}}
\frac{p_{\gamma }}{4},\quad P_-=\bar P_+,\quad
P_3=\frac{p_\gamma}{2}-{\bf i}\left(z\pi-\bar
z\bar\pi\right).\label{ier}\ee In the same way we introduce the
Hamiltonian generators $I_a$ corresponding to the vector fields
(\ref{vec2}): \be I_3=-\frac{p_\gamma}{2},\quad I_+=\frac{I_2-{\bf
i} I_1}{2}=\frac{{\bf i} p_\gamma z+2\bar \pi\left(1+z \bar
z\right)}{4}e^{-2{\bf i} \gamma},\quad I_-=\overline{I}_+
\label{per}.\ee

These quantities define, with respect to the Poisson bracket
(\ref{pb}), the $so(4)=so(3)\times so(3)$ algebra \be
\left\{P_i,P_j\right\}=\varepsilon_{ijk}P_c,\quad\left\{I_i,I_j\right\}=\varepsilon_{ijk}I_k,
\quad\left\{I_i,P_j\right\}=0. \label{pbso4}\ee The functions
$P_i$, $I_i$ obey the following equality, important for our
considerations \be I_k I_k=P_k P_k. \label{IP}\ee At this point we
replace the initial Lagrangian (\ref{lag0}) by the following one,
which is variationally equivalent
 \be {\cal
L}_{int}=2\left(P_+\Lambda_++P_-\Lambda_-+P_3\Lambda_3\right)-P_iA_i-\frac{P_iP_i}{g
r}-\frac{g r A_iA_i}{4} +g\left({\dot r}_+{\dot r}_- +{\dot
r}^2_1\right). \label{lag7}\ee The isometries of this modified
Lagrangian  corresponding to (\ref{vec2}) are defined by the vector
fields \be {\bf\widetilde U }_i\equiv \{I_i,\;\;\}, \label{is1}\ee
where $I_i$ are given by (\ref{per}) and the Poisson brackets are
given by (\ref{pb}). The quantities $I_i$ entering (\ref{is1}) are
the Noether constants of motion of the modified Lagrangian
(\ref{lag7}). This can be easily seen taking into account the
following equality
 \be
2\left(P_+\Lambda_++P_-\Lambda_-+P_3\Lambda_3\right)=p_\gamma\dot{\gamma}+\pi\dot{z}+\bar\pi\dot{\bar
z}.\label{pp}\ee We have now to perform the reduction via the action
of the $SU(2)$ group given by the vector fields (\ref{is1}). For
this purpose we have to fix the Noether constants of motion
(\ref{per}), setting
 $$I_k=const,\qquad I_kI_k\equiv s^2 .$$
 Since  the constants of
motion $I_k$ do not dependent on the $r_\pm, r_5$ coordinates we can
perform an orthogonal rotation so that only the third component of
this se, $I_3$, assumes a value different from zero. Equating $I_+$
and $I_-$ with zero we obtain: \be -I_3=\frac{p_\gamma}{2}=s,\qquad
\bar \pi= {{\bf i}}s \frac{z}{1+z\bar z}, \quad \pi= -{{\bf i}}s
\frac{\bar z}{1+z\bar z}.\quad  \ee Hence,
 \be
P_+=-{\bf i}s\frac{\bar z}{1+z \bar{z}},\quad P_-={\bf
i}s\frac{z}{1+z \bar{z}},\quad P_3=-s\frac{1-z\bar z}{1+z\bar
z}\label{fix}.\ee Therefore $P_k$ coincide with the Killing
potentials of the ${ S}^2$ sphere! This is by no means an occasional
coincidence.

Taking in mind the equality (\ref{pp}) we can conclude that the
third term entering (\ref{lag7}) can be ignored because it is a
full time derivative. Besides that, taking into account
(\ref{IP}), we can rewrite the Lagrangian as follows:
 \be{\cal
L}_{red}=\frac{\widetilde{g}\dot{x}_{\mu}\dot{x}_{\mu}}{2}- {\bf i}
s\frac{\bar {z}\dot{z}-z \dot{\bar z}}{1+z\bar z} -{sh_k(z,\bar
z)}A_k-\frac{s^2}{2r^2\widetilde{g}},\qquad
{\widetilde{g}}\equiv\frac{g}{2r}, \qquad \mu=1,\ldots,5, \ee where
we have used the identity $$-\frac 14 g r A_iA_i +g\left({\dot
r}_+{\dot r}_- +{\dot r}^2_1\right)= g\frac{{\dot{x}}_\mu
{\dot{x}}_\mu}{4r}.$$
 The second term in the above
reduced Hamiltonian is the one-form defining the symplectic (and
K\"ahler) structure on ${  S}^2$, while $h_k$ given in (\ref{kil}) are the Killing potentials defining the isometries of the
K\"ahler structure. We have in this way obtained the Lagrangian describing the motion of a five-dimensional isospin particle
in the field of an $SU(2)$ Yang monopole. The metric of the configuration space is defined
by the expressions ${\widetilde g}_{\mu\nu}=
\frac{g}{2r}\delta_{\mu\nu}$. For a detailed description of
the dynamics of the isospin particle we refer to  \cite{horv}.

Similarly to the $U(1)$ case, the reduced system is specified by the presence
of a centrifugal potential $s^2/2{\widetilde g}r^2$, which essentially cancels the impact of the monopole
in the classical and quantum solutions of the system. Particularly, for spherically symmetric systems
(including those  with extra potential terms),
the impact of the Yang monopole on the spectrum implies a change in the validity range of the orbital momentum \cite{mny}.
In
supersymmetric systems, on the other hand, the presence of a
monopole can change essentially the supersymmetric properties.

It therefore follows that the Noether constants of motion do not
allow us to exclude the $z,\bar z$ variables. However, their time
derivatives appear in the Lagrangian in a linear way only and
define the internal degrees of freedom of the five-dimensional
isospin particle interacting with a Yang monopole. As a
consequence, the dimensionality of the phase space of the reduced
system is $2\cdot 5+2=12$. Only  for  the particular case $s=0$,
corresponding to the absence of the Yang monopole, we obtain a
five-dimensional system. This means that  {\sl locally} the
Lagrangian of the system can be formulated in a six-dimensional
space. Such a representation seems, however, useless, in contrast
with the one presented here.\par
The further reduction of the constructed $(5+...)$-dimensional
system to a $(4+...)$-dimensional one would be completely similar
to the $U(1)$ case: it requires fixing the radial variable $r$.
The resulted system describes the isospin particle moving in a
four-dimensional space and interacting with the BPST
instanton. \par
In this Section we have considered the Lagrangian reduction
procedures, restricting ourselves to $2p$-dimensional systems with
{\sl conformal flat metrics} only. From  our considerations it
is however clear that similar reductions can be performed also for
particles moving on other $G$-invariant $2p$-dimensional spaces
(not necessarily conformally flat), in presence of a $G$-invariant
potential. The modifications do not yield any qualitative
difference with the proposed reduction procedures and will be
reflected in more complicated forms of the resulting Lagrangians.
The possibility of adding to the initial system $G$-invariant potentials is obvious.

\setcounter{equation}{0}
\section{Supersymmetry}

We discuss now the supersymmetric extensions, both for $p=2$ and $p=4$, of the bosonic constructions we have
dealt so far. For our purposes we have to ensure the compatibility of the supersymmetry transformations
acting on the ``root'' or ``minimal length'' supermultiplets $(2p,2p,0)$, with the bilinear transformations
\be\label{bilin}
x_\mu = u^T\gamma_{\mu}u,
\ee
where, for $p=2$, $\mu=1,2,3$ and the $\gamma_{\mu}$'s are the generators of the Euclidean Clifford algebra $Cl(3,0)$ while,
for $p=4$, $\mu=1,2,3,4,5$,
the $\gamma_\mu$'s are the generators of the Euclidean Clifford algebra $Cl(5,0)$.

In the $p=2$ case we can choose
\be\label{gamma30}
\gamma_1 = {\bf 1}_2\otimes {\tau_1},\quad
\gamma_2 = {\bf 1}_2\otimes {\tau}_2, \quad
\gamma_3 = \tau_A\otimes \tau_A,
\ee
where
\be\label{tau} \tau_1= \left( {\begin{array}{*{20}c}
   0 & 1  \\
   1 & 0  \\
\end{array}} \right),\quad
 \tau_2= \left( {\begin{array}{*{20}c}
   1 & 0  \\
   0 & -1  \\
\end{array}} \right),\quad
\tau_A= \left( {\begin{array}{*{20}c}
   0 & 1  \\
   -1 & 0  \\
\end{array}} \right)
\quad {\bf 1 }_2= \left( {\begin{array}{*{20}c}
   1 & 0  \\
   0 & 1  \\
\end{array}} \right).
\ee
Due to the Schur's lemma \cite{oku}, the three gamma matrices in (\ref{gamma30}) commute with a single
matrix
\be\label{sigma3}
\sigma_3= {\tau_A}\otimes {\bf 1}_2
\ee
(${\sigma_3}^2={\bf 1}_4$) which defines the complex structure in $Cl(3,0)$.\par

For the $p=5$ case the $\gamma$-matrices look as follows
\be\label{gamma50}
\begin{array}{c}
\gamma_1=\tau_A\otimes\tau_1\otimes\tau_A,\\
\gamma_2=\tau_A\otimes\tau_2\otimes\tau_A,\\
\gamma_3=\tau_A\otimes\tau_A\otimes{\bf 1}_2,\\
\gamma_4=\tau_1\otimes{\bf 1}_2\otimes{\bf 1}_2,\\
\gamma_5=\tau_2\otimes{\bf 1}_2\otimes{\bf 1}_2,
\end{array}.
\ee
where the matrices $\tau_1,\tau_2,\tau_A$ are defined in (\ref{tau}).

The real coordinates $u_{a}$, $a=1,\ldots 2p$,
are related with the complex/quaternionic coordinates ${\bf u}_\alpha, {\bf\overline{ u}}_\alpha$ considered in the previous
Sections, by the expressions
 \be {\bf u}_1=u_{4 }+ {\bf
e}_iu_{i},\qquad {\bf u}_2=u_{8 }+ {\bf
e}_iu_{4+i},\qquad i=1,2,3. \ee
The ${\bf V}=\partial\varphi$ vector field defining, in the $p=2$case,  the $U(1)$ isometry, therefore looks
\be
{\bf V}=u^T\sigma_3\frac{\partial}{\partial u}.
\ee
In the $p=4$ case, the ${\bf U}_i$ vector fields defining the  ${SU(2)}$ isometries are given by the expressions \be {\bf
U}_i=u\Sigma_i\frac{\partial}{\partial u}, \qquad \Sigma_1 ={\bf
1}_2\otimes\tau_A\otimes\tau_1,\quad \Sigma_2 ={\bf
1}_2\otimes\tau_A\otimes\tau_2,\quad \Sigma_3 ={\bf
1}_2\otimes{\bf1}_2\otimes\tau_A. \label{sigma}
\ee
It is easily
proven that the $su(2)$ matrix generators $\Sigma_i$ commute with
the Gamma-matrices $\Gamma_\mu$ ($[\Sigma_i,\Gamma_\mu]=0 $). This
is in agreement with the fact that ${\bf U}_i$ define the
isometries of the eight-dimensional Lagrangian (\ref{lag0}).
\par

The relation pointed out in \cite{pashtop} between Clifford algebra and the associated supersymmetric root
multiplets has a consequence that the Schur's lemma induces real, complex or quaternionic structures, see
\cite{kt} and \cite{krt}, on the minimal length multiplets.\par
For $p=2$, the $(4,4,0)$ root multiplet is an ${\cal N}=4$ quaternionic multiplet, since the supersymmetry algebra
\be
{Q}_a{Q}_b+{Q}_b{Q}_a=
\delta_{ab}{\bf 1}{\partial_t},\qquad {Q}_a {H}-{H}{Q}_a=0,\qquad\qquad
{H}\equiv {\bf 1}\partial_t,\quad a,b=1,\ldots,{\cal N}=4
\label{susy0}\ee
is realized through the supermatrices acting on the $(u_1,u_2,u_3,u_4;\psi_1,\psi_2,\psi_3,\psi_4)$ multiplet, given by
\be\label{n444susy}
{Q}_{4}=\left(
\begin{array}{cc}
0&{\bf1}_{4}\\
{\bf 1}_{4}{\partial_t}& 0
\end{array}\right),\qquad
{Q}_{i}=\left(
\begin{array}{cc}
0&{\widehat\gamma}_{i}\\
-{\widehat\gamma}_{i}\partial_t & 0
\end{array}\right),\qquad {i}=1,2,3,
\label{qqq}\ee
where
\be\label{gamma03}
{\widehat\gamma}_1 = {\tau}_A\otimes \tau_1,\quad
{\widehat\gamma}_2 = \tau_A\otimes {\tau}_2, \quad
{\widehat\gamma}_3 = {\bf 1}_2\otimes \tau_A
\ee
and $Q_i$, $Q_4$ all commute with the three matrices ${\widetilde \Sigma}_j =\sigma_j\otimes\sigma_j$, $j=1,2,3$,
($\sigma_1=\tau_1\otimes\tau_A$, $\sigma_2=\tau_2\otimes \tau_A$, while $\sigma_3$
is given by (\ref{sigma3})).
Notice that ${\widetilde \Sigma}_1, {\widetilde \Sigma}_2$ (contrary to ${\widetilde \Sigma}_3$) do not
leave invariant the coordinates $x_1,x_2,x_3$ entering, for $p=2$,
(\ref{bilin}).
\par
For $p=4$ the situation is as follows. According to the supersymmetric extension of the Schur's lemma,
\cite{krt} and \cite{kt2}, there are at most ${\cal N}=5$ supersymmetry generators commuting with the $su(2)$
generators ${\widetilde \Sigma}_j$ (now ${\widetilde \Sigma}_j =\Sigma_j\oplus\Sigma_j$, with $\Sigma_j$ given
 in (\ref{sigma})) and acting on the $(8,8,0)$ root multiplet\footnote{An extra pseudosupersymmetry operator,
${\widetilde Q}$, such that ${\widetilde Q}^2=-H$, is allowed.}.
\par
The ${\cal N}=8$ supersymmetry transformations acting on the root multiplet with fields $(u_a; \psi_b)$, ($a,b=1,2,\ldots, 8$)
 are given by
\be\label{N8susies}
Q_k =\left(\begin{array}{cc}
0 & \gamma_k \\
-\gamma_k\cdot H & 0 \\
\end{array}
\right), \qquad
Q_8 = \left(
\begin{array}{cc}
0 & {\bf 1}_8 \\
{\bf 1}_8\cdot H & 0 \\
\end{array}
\right),\qquad k=1,2,\ldots, 7,
\ee
where
\be
\begin{array}{llll}
\gamma_1 = \tau_1\otimes \tau_A\otimes {\bf 1}_2,&
\gamma_2 =\tau_2\otimes \tau_A\otimes {\bf 1}_2,&
\gamma_3 =\tau_A\otimes {\bf 1}_2\otimes \tau_1,&
\gamma_4 = \tau_A\otimes {\bf 1}_2\otimes \tau_2,
\\
\gamma_5 = {\bf 1}_2\otimes \tau_1\otimes \tau_A,&
\gamma_6 = {\bf 1}_2\otimes \tau_2\otimes \tau_A,&
\gamma_7 = \tau_A\otimes \tau_A\otimes \tau_A.&
\end{array}
\ee

The subset of ${\cal N}=5$ supersymmetry transformations commuting with the above specified $su(2)$ generators
${\widetilde \Sigma}_j$,
\be
\relax [Q_I,{\widetilde \Sigma}_j]=0,
\ee
is explicitly given by
$Q_1, Q_2, Q_5, Q_6, Q_8$.\par
In accordance with the above results, the reduced Lagrangians, invariant under the extended supersymmetry
algebra and compatible with the $G$-group action, where $G=U(1)$ for $p=2$ and $G=SU(2)$ for $p=4$,
can be recast in a complex and, respectively, quaternionic formalism. We will discuss them separately in the next
subsections.

\subsection{The $U(1)$ reduction}

We discuss the reduction of the ${\cal N}=4$ supersymmetric systems with a $({4,4,0})$ supermultiplet.

The three $U(1)$-invariant fields $x_1,x_2,x_3$ constructed in (\ref{bilin}), for $p=2$, as bilinear combinations of the four $u_i$ fields,
transform under ${\cal N}=4$ supersymmetry with transformations induced by (\ref{n444susy}). It is easily verified that the induced
supersymmetry closes linearly and the resulting supermultiplet corresponds to the ${\cal N}=4$ $(3,4,1)$ fields content where, in addition
to the three $x_\mu$, we have $4$ fermions and an auxiliary bosonic field. All the fields belonging to this multiplet
are $U(1)$-invariant and given by bilinear combinations of the $u_i$ and $\psi_i$ fields
entering the original $(4,4,0)$ supermultiplet. \par
The commutativity of the ${\cal N}=4$ supersymmetry algebra with the $U(1)$ generator makes possible to use an alternative description,
more suitable in describing the ${\cal N}=4$ supersymmetric quantum mechanical system in presence of a monopole. It makes use of the
complex coordinates (bosonic and, respectively, fermionic)  ${\bf u}_\alpha$, ${\psi_\alpha}$ and the
``chiral supercharge" generators $Q^{\pm}_k =Q_k\pm\imath Q_{k+2}$ ($k=1,2$).
The supersymmetry transformations can therefore be re-expressed as
\be\begin{array}{llll}
Q^{+}_1{\bf u}_\alpha=\psi_\alpha, & Q^{+}_1 \psi_\alpha={\bf\dot u}_\alpha,&
Q^{-}_1 {\bf\overline u}_\alpha={\overline \psi}_\alpha, & Q^{-}_1 {\overline \psi}_\alpha=
{\bf\dot{\overline  u}}_\alpha,\\
 Q^{+}_2 {\bf u}_\alpha=\epsilon_{\alpha\beta}\psi_\beta, & Q^{+}_2 \psi_\alpha=\epsilon_{\alpha\beta}{\bf\dot u}_\beta, &
 Q^{-}_2 {\bf\overline u}_\alpha=\epsilon_{\alpha\beta}{\overline \psi}_\beta, & Q^{-}_2 {\overline \psi}_\alpha=\epsilon_{\alpha\beta}{\bf\dot{\overline  u}}_\beta,\\
 Q^{+}_k {\bf\overline u}_\alpha=0, & Q^{+}_k {\overline \psi}_\alpha=0.&&
\end{array}
\label{susy4}\ee
The $U(1)$ group acts on the complex variables $({\bf u}_\alpha, \psi_{\alpha})$ as follows
\be
 {\bf u}_\alpha\to {\rm e}^{\imath\kappa} {\bf u}_\alpha,
\quad {\bf\overline u}_\alpha\to {\rm e}^{-\imath\kappa}{\bf\overline u}_\alpha ,
 \quad \psi_\alpha\to {\rm e}^{\imath\kappa} \psi_\alpha,
 \quad {\overline \psi}_\alpha\to {\rm e}^{-\imath\kappa}{\overline \psi}_\alpha,
 \ee
 where $\kappa$ is arbitrary real parameter.
\par
By reducing the ${\cal N}=4$ $(4,4,0)$ supersymmetric system above via the $U(1)$ group action we obtain a system still possessing
the ${\cal N}=4$ supersymmetry.
This is reached by choosing, in complete analogy with the bosonic case, besides the three  $U(1)$-invariant
 bosonic coordinates (\ref{bilin}),
four $U(1)$ invariant fermionic coordinates $\chi_\alpha$ given below and
an extra-bosonic field $2\varphi=\imath\log {\bf u}_1/{\bf\overline u}_1$.
The whole set of coordinates of the reduced system are \cite{Ivanov:2003nk}
 \be\label{newn4}
 {\bf x}=2{\bf\bar u}_1 {\bf u}_2,\quad x_3={\bf\bar u}_1 {\bf u}_1-{\bf\bar u}_2 {\bf u}_2,\quad
 \chi_\alpha={\rm e}^{-\imath\phi}\psi_\alpha,\quad {\overline\chi}_\alpha={\rm e}^{\imath\phi}{\overline\psi}_\alpha.
 \ee
The general ${\cal N}=4$ Lagrangian constructed with the $({4,4,0})$ supermultiplet is given by
(see, e.g., \cite{root})\footnote{A supersymmetric hamiltonian in presence of an $U(1)$ monopole was first constructed in \cite{sm1}
based on the construction \cite{CR} of supersymmetric quantum mechanical systems from dimensional reduction of higher dimensional
superfield theories. In \cite{sm1} the reduction of the chiral supersymmetric QED was considered. The lowest order effective action
produces a supersymmetric sigma-model with constant metric while, when the Born-Oppenheimer corrections become large, a non-trivial
metric is recovered \cite{sm2}.}.
\be
{\cal L}_4^{SUSY}={\cal L}_4 +\frac{\imath g(u,\overline u)}{2}\left({\overline\psi}\cdot{D_t\psi}-
{D_t{\overline\psi}}\cdot{\psi}\right) -
{\cal R}(\psi\cdot {\overline\psi})(\psi\cdot {\overline\psi}),\quad D_t\psi\equiv{\dot\psi}+\Gamma\psi{\dot u},
\ee
where $D\psi$ is defined by the connection of the  metric
$ds^2=gdu\cdot d{\overline u}$,  ${\cal R}$ is the curvature of this connection
 and ${\cal L}_4$ is the bosonic Lagrangian given in (\ref{lag0}).
 Therefore, for a $U(1)$ invariant metric, the supersymmetric Lagrangian also possesses an $U(1)$ invariance.

When re-writing the initial system in terms of ${\bf r_\pm}, r_1, \chi_\alpha,{\overline\chi}_\alpha,\varphi$,
we recover that $\varphi$ is a cyclic variable. Excluding it, in analogy with the bosonic case, we obtain
an ${\cal N}=4$ supersymmetric system with $3$ bosonic dimensions.
The presence of the fermionic degrees of freedom does not yield qualitative changes in the reduction procedure.
The bosonic reduction procedure discussed in Subsection 3.1 is consistently implemented in the supersymmetric case as well.

\subsection{The $SU(2)$ reduction}

We discuss now the reductions of the $({8,8,0})$ supersymmetric multiplet
via the $SU(2)$ group action.
In contrast with the previous case, the $su(2)$ algebra does not commute which the whole set of the ${\cal N}=8$ supersymmetry
generators (\ref{N8susies}). For that reason the reduced system cannot inherit the whole ${\cal N}=8$ supersymmetry, but only
its ${\cal N}=5$ subalgebra (we recall that an explicit presentation of the supersymmetry tranformations is given by
$Q_1, Q_2, Q_5, Q_6,Q_8$ entering (\ref{N8susies})). \par
It is worth mentioning that
there are ${\cal N}=6$ supersymmetry generators commuting with the $U(1)$ group action defined, e.g., by ${\widetilde\Sigma}_3$
(the extra supersymmetry generator closing ${\cal N}=6$ corresponds to $Q_7$).
As a consequence, the $U(1)$ reduction of the $({8}, {8})$ supermultiplet produces an ${\cal N}=6 $ supersymmetric
mechanics on $C P^3$ in presence of a constant magnetic field. The reduction by the whole $SU(2)$ group yields further restrictions
on the number of  supersymmetries since at most
${\cal N}=5$ supersymmetry generators commute with the $su(2)$ generators which define the quaternionic structure.

In order to exploit the quaternionic properties it is convenient to redefine the $(8,8)$ variables as follows
\be
\begin{array}{llllllll}
u_1\to v_0, &u_2\to v_2,&   u_3\to v_3,& u_4\to v_1,& u_5\to {\bar v}_0,& u_6\to {\bar v}_2,
& u_7\to \bar v_3,& u_8\to {\bar v}_1\\
\psi_1\to\bar\lambda_0,&
\psi_2\to\bar\lambda_2,&\psi_3\to\bar\lambda_3,&\psi_4\to\bar\lambda_1,&\psi_5\to\lambda_0,&\psi_6\to\lambda_2,&
\psi_7\to\lambda_3,&\psi_8\to\lambda_1.
\end{array}
\label{redef}\ee
After this redefinition the ${\cal N}=5$ supersymmetry transformations take the following form.\\
The $Q_i$ ($i=1,2,3$) transformations are ($\epsilon_{123}=1$):
\be\begin{array}{llll}
Q_iv_0=\lambda_i,&
Q_iv_j = -(\delta_{ij}\lambda_0+\epsilon_{ijk} \lambda_k),&
Q_i{\overline v}_0= -{\overline\lambda}_i,&
Q_i{\overline v}_j = \delta_{ij}{\overline\lambda}_0+\epsilon_{ijk} {\overline\lambda}_k,\\
Q_i{\lambda}_0= -{\dot v}_i,&
Q_i{\lambda}_j = \delta_{ij}{\dot v}_0+\epsilon_{ijk} {\dot v}_k,&
Q_i{\overline \lambda}_0= {\dot {\overline v}}_i,&
Q_i{\overline \lambda}_j = -(\delta_{ij}{\dot{\overline v}}_0+\epsilon_{ijk} {\dot{\overline v}}_k).
\end{array}\label{qitransf}
\ee
The $Q_4$ transformation is
\be\begin{array}{llll}
Q_4v_0= \lambda_0,&
Q_4v_j = \lambda_j,&
Q_4{\overline v}_0= {\overline\lambda}_0,&
Q_4{\overline v}_j = {\overline\lambda}_j,\\
Q_4{\lambda}_0= {\dot v}_0,&
Q_4{\lambda}_j = {\dot v}_j,&
Q_4{\overline \lambda}_0= {\dot {\overline v}}_0,&
Q_4{\overline \lambda}_j = {\dot{\overline v}}_j.\end{array}\label{q4transf}
\ee
The $Q_5$ transformation is
\be\begin{array}{llll}
Q_5v_0= {\overline\lambda}_0,&
Q_5v_j = {\overline \lambda}_j,&
Q_5{\overline v}_0= -{\lambda}_0,&
Q_5{\overline v}_j = -{\lambda}_j,\\
Q_5{\lambda}_0= -{\dot{\overline v}}_0,&
Q_5{\lambda}_j = -{\dot{\overline v}}_j,&
Q_5{\overline \lambda}_0= {\dot {v}}_0,&
Q_5{\overline \lambda}_j = {\dot{v}}_j.
\end{array}\label{q5transf}
\ee
The ${\widetilde Q}$ pseudosupersymmetry operator (${\widetilde Q}^2=-H$) which commutes
with the $su(2)$ generators is given by
\be\begin{array}{llll}
{\widetilde Q}v_0={\overline\lambda}_0,&
{\widetilde Q}v_j = {\overline \lambda}_j,&
{\widetilde Q}{\overline v}_0= {\lambda}_0,&
{\widetilde Q}{\overline v}_j = {\lambda}_j,\\
{\widetilde Q}{\lambda}_0= -{\dot{\overline v}}_0,&
{\widetilde Q}{\lambda}_j = -{\dot{\overline v}}_j,&
{\widetilde Q}{\overline \lambda}_0= -{\dot {v}}_0,&
{\widetilde Q}{\overline \lambda}_j = -{\dot{v}}_j.
\end{array}\label{qpstransf}
\ee
Notice that the pseudosupersymmetry operator ${\widetilde Q}$, together with $Q_5$, can be used to define
a BRST-type transformation $Q_{BRST}$ ($Q_{BRST}^2=0$) given by $Q_{BRST}=\frac{1}{2} (Q_5+{\widetilde Q})$, such that
\be\begin{array}{llll}
Q_{BRST}v_0= {\overline\lambda}_0,&
Q_{BRST}v_j = {\overline \lambda}_j,&
Q_{BRST}{\overline v}_0= 0,&
Q_{BRST}{\overline v}_j = 0,\\
Q_{BRST}{\lambda}_0= -{\dot{\overline v}}_0,&
Q_{BRST}{\lambda}_j = -{\dot{\overline v}}_j,&
Q_{BRST}{\overline \lambda}_0= 0,&
Q_{BRST}{\overline \lambda}_j = 0.
\end{array}\label{qbrsttransf}
\ee
The BRST-operator $Q_{BRST}$ commutes with the $su(2)$ generators and anticommutes
with the remaining ${\cal N}=4$ $su(2)$-invariant supercharges.
\par
The most general $su(2)$-invariant ${\cal N}=4,5$ actions for the $(8,8)$ multiplet can be
 computed with the construction of \cite{krt} (further developed in \cite{grt}). A manifestly ${\cal N}=4$ invariant action
is obtained from the lagrangian
  \be
{\cal L}=Q_1Q_2Q_3Q_4 f(v, \bar v),
\ee
where  the supercharges $Q_1,\ldots,Q_4$ are given by (\ref{qitransf}), (\ref{q4transf}) and $f$
is an {\em unconstrained} function of the bosonic coordinates $v_0,v_1,v_2,v_3,{\bar v}_0,{\bar v}_1, {\bar v}_2,
{\bar v}_3$. The explicit expression for ${\cal L}$, obtained with the help of a package for Maple 11 and written in terms
of the quaternionic structure constants, is reported for completeness in the Appendix. \par
The ${\cal N}=5$ invariance is obtained by a constraint, induced by the fifth $su(2)$-invariant supersymmetry transformation
$Q_5$, which requires $Q_5{\cal L}$ be a total time-derivative. The ${\cal N}=5$ requirement implies that $f$ must satisfy the equation
\be\label{n5constraint}
\Delta_8 f\equiv f_{\mu\mu}+f_{{\bar\mu}{\bar\mu}}=0,
\ee
where $\mu=0,1,2,3$ and $f_\mu\equiv\partial f/\partial v_\mu$, $f_{\bar\mu}\equiv\partial f/\partial {\bar v}_{\mu}$.
\par
An alternative constraint is obtained by requiring both the ${\cal N}=4$ invariance {\em and} the $Q_{BRST}$ invariance.
In this case $f$ must satisfy
\be
\Delta_4 f\equiv f_{\mu\mu}=0.
\ee
In order to have an $su(2)$-invariant action, an $su(2)$-invariant constraint has to be imposed on $f$.
This constraint can be explicitly solved by expressing $f$ not directly in terms of $v_\mu$, ${\overline v}_\mu$ (or $u_1,\ldots, u_8$),
but through the $su(2)$-invariant ``bilinear coordinates'' $x_\mu$ (now $\mu=1,2,3,4,5$) entering (\ref{bilin}). We obtain as a
result an $su(2)$-invariant, ${\cal N}=5$ supersymmetric lagrangian for a $5$-dimensional system (given by the $x_\mu$ coordinates).\par
In analogy with the case discussed in the previous subsection, we can compute the supermultiplet generated by the $5$ $su(2)$-invariant
bilinear fields $x_\mu$. Its fields content is given \cite{kt2} by $(5,11,10,5,1)$. This supermultiplet corresponds
to a $(1,5,10,10,5,1)\rightarrow (0,5,11,10,5,1)$ dressing of the ${\cal N}=5$
``enveloping multiplet'' (see \cite{krt}), whose fields content is given by Netwon's binomials.
All fields entering $(5,11,10,5,1)$ are $su(2)$-invariant and given by bilinear combinations of the original $u_i,\psi_i$ fields.
This multiplet contains twice as many fields entering a minimal (irreducible,
in physicists' language) ${\cal N}=5$ multiplet. It is a reducible, but indecomposable, multiplet
which can be better described in the basis of the irreducible $(5,8,3,0,0)$ and $(0,3,5,5,1)$ (see \cite{krt}) supermultiplets.
 Just as in the previous case, the ${\cal N}=5$ supersymmetry is realized linearly on $(5,11,10,5,1)$.
 It is worth pointing out that, of course, we are not in presence of a doubling of the degrees of freedom. The $(5,11,10,5,1)$
 multiplet consists of composite fields (bilinear combinations of the original fields). It has been observed before
 (see e.g. in \cite{krt} the discussion of the tensor product of the ${\cal N}=4$ $(1,4,3)$ multiplet) the phenomenon
 of a composite multiplet whose number of component fields is twice as many the number of the generating fields
 expressing its composite fields. No contradiction arises. $(5,11,10,5,1)$ carries a linear representation.
Its component fields, however, can be expressed as composite fields of a ``smaller'' multiplet.
 \par
An important comment has to be made. In \cite{grt} it has been explicitly proven that requiring the
${\cal N}=5$ invariance for an off-shell action based on the $(2,8,6)$ multiplet, automatically induces
a full ${\cal N}=8$ invariance. Similarly, the ${\cal N}=5$ invariance constraint (\ref{n5constraint})
for the $(8,8,0)$ multiplet automatically guarantees an ${\cal N}=8$ invariance.
This is in agreement with the result of the first paper in \cite{k1}, where the same constraint was derived by requiring the
whole ${\cal N}=8$ invariance, and with \cite{Ivanov:2007sh}, where the general  superfield and
component actions of this multiplet were explicitly given. It was also proven there that the $8$-dimensional  harmonicity
condition for the Lagrangian is
a necessary and sufficient condition to have an ${\cal N}=8$ supersymmetry.
 Therefore, combining (\ref{n5constraint}) and the $SU(2)$ constraint
(expressed by the fact that $f$ is function of the five bilinear coordinates entering (\ref{bilin}))
produces an ${\cal N}=8$ $SU(2)$-invariant system. On the other hand, the three extra supersymmetry
generators (the ones which do not commute with the $su(2)$ algebra generators) are not essential to
derive the symmetries of the action. They also close on a much larger multiplet than $(5,11,10,5,1)$,
containing fields which are not $SU(2)$-invariant. We recall that the $SU(2)$ group acts on the
 fields entering $(5,11,10,5,1)$ as the identity operator. Furthermore, a quaternionic structure is
 only available for the ${\cal N}=5$ subalgebra.

The supersymmetry transformations (\ref{qitransf}), (\ref{q4transf}), (\ref{q5transf})
preserve the quaternionic structure. We can therefore express the ${\cal N}=5$ $(8,8,0)$ component fields in
a quaternionic framework,
in such a way that the $SU(2)$ group action is expressed through
\be
{\bf u}_\alpha\to {\bf G} {\bf u}_\alpha , \Psi_\alpha\to {\bf G}\Psi_\alpha,\quad
{\rm where}\quad {\bf G}{\bar {\bf G}}=1, \quad {\bf G},\; {\bf u}_\alpha ,\;\Psi_\alpha\in\DH,\qquad \alpha=1,2.
\ee
In this language the $5$ bilinear coordinates $x_\mu$ and $8$ $SU(2)$-invariant
fermions can be expressed as follows
\be\label{newn5basis}
{\bf x}=2{\bf\bar u}_1{\bf u}_2, \quad x_5={\bf\bar u}_1{\bf u}_1 -{\bf\bar u}_2{\bf u}_2,
  \quad {\bf \chi}_\alpha={\bf \bar g}\Psi_\alpha.
\ee

These positions mimic, in the $SU(2)$ reduction case, what happens in the $U(1)$ case. They suggest the existence of a
supersymmetric description of a five-dimensional system with a Yang monopole realizing
the ${\cal N}=5$ supersymmetry non-linearly on a $(5,8,3)$ field content. The main difference with respect to
the $U(1)$ case is the fact that the supersymmetric $SU(2)$-invariant multiplet realized with bilinear combinations of
the $(8,8,0)$ fields contain twice as many fields as the ones entering $(5,8,3)$. A possible strategy consists in
extracting the linear $(5,8,3)$ multiplet entering $(5,11,10,5,1)$ by setting equal to zero the fields entering its $(0,3, 7,5,1)$
submultiplet. A non-linear transformation allows to re-express
the $(5,8,3)$ fields entering the bilinear basis with the $(5,8,3)$ fields entering (\ref{newn5basis}).
This issue will be detailed in a forthcoming publication.

\par
\par

\section{Summary and Discussion}

Let us briefly summarize our results.
We investigated the properties of the supersymmetric mechanics associated with the second Hopf map.
We found that the reduction via the $SU(2)$ group action of the $(8,8,0)$ multiplet generates a
five-dimensional supersymmetric multiplet induced by the ${\cal N}=5$ supersymmetry generators acting on $(8,8,0)$
and commuting with the $su(2)$ algebra generators. The resulting supermultiplet is a reducible, but indecomposable
length-$5$ multiplet with fields content $(5,11,10,5,1)$. The $SU(2)$ action on this field coincides with the identity operator.
The resulting invariant action
has been explicitly computed. We proved that it admits both $SU(2)$ invariance and an ${\cal N}=8$ invariance.
The invariance under the three extra supersymmetry operators is less important for two reasons. The first one is that
it is automatically induced by the invariance under the ${\cal N}=5$ $SU(2)$-invariant operators. The second one is that
the $N=8$ action closes on a much larger multiplet than $(5,11,10,5,1)$ and the extra-fields are inessential to derive
the invariant action.

We further pointed out that an  extra, BRST-like, symmetry can be imposed on
the reduced system. Constraining the $5$ $SU(2)$-invariant coordinates living on the surface of the $ S^4\subset \DR^5$ sphere,
produces an ${\cal N}=5$ non-linear multiplet generated by the $4$ angular coordinates
of the sphere. The description of a system in presence of an $SU(2)$ Yang monopole/BPST instanton requires further work.
At first the $(5,11,10,5,1)$ linear multiplet should be decomposed into its two basic irreducible constituents $(5,8,3,0,0)$
and $(0,3,5,7,1)$ (the latter is a length-$4$ ${\cal N}=5$ multiplet
first described in \cite{krt}); next the fields entering the $(0,3,5,7,1)$ multiplet should be consistently set to zero.
As we discussed in the previous section, the `´doubling'' of the fields entering the $(5,11,10,5,1)$ multiplet is a
reflection of the composite nature of its component fields.
The fields entering $(5,8,3,0,0)$ can now be equated, through non-linear transformations, with the $(5,8,3)$ $SU(2)$-invariant
fields describing the Yang monopole and introduced in (\ref{newn5basis}). Due to the non-linearity of the transformation,
the ${\cal N}=5$ supersymmetry is realized non-linearly in this new basis. This procedure corresponds to its simpler $U(1)$
counterpart concerning the reduction of ${\cal N}=4$
$(4,4,0)$ into ${\cal N}=4$ $(3,4,1)$. It is worth pointing out that,
in contrast with the $U(1)$ reduction case, for the non-abelian $SU(2)$ reduction the auxiliary fields cannot
be completely removed from the Lagrangian. Indeed, they ``partially'' transmute into isospin degrees of freedom. This difference
between the two reduction procedures was expected from the beginning, since it has   a purely bosonic origin. Much less expected
are the subtle issues concerning the supersymmetric reductions. For the $U(1)$ reduction, the whole set of ${\cal N}=4$ extended
supersymmetries is $U(1)$ invariant while, for $SU(2)$, only  ${\cal N}=4$ or ${\cal N}=5$ of the original ${\cal N}=8$ supersymmetries
are $SU(2)$-invariant.\par
As remarked in the previous Section, by making the reduction with respect to the $U(1)$ group for the ${\cal N}=8$ $(8,8)$ supermultiplet
we obtain a supersymmetric quantum mechanics on $CP^3$ in presence of a constant magnetic field and with ${\cal N}=6$ supercharges
commuting with the $u(1)$ algebra generator.\par
In this work we prepared the ground for further developments, clarifying the general features of the supersymmetric reductions
and postponing to forthcoming papers the detailed descriptions.
\par~
\par

 \renewcommand{\theequation}{A.\arabic{equation}}
  \setcounter{equation}{0}  
  \section*{Appendix}  

For completeness we are reporting the ${\cal N}=4$ $su(2)$-invariant lagrangian ${\cal L}$ for the $(8,8)$ multiplet,
expressed in terms of the quaternionic structure constants. After setting $\epsilon_{123}=+1$,
$\Gamma=f_{00}+f_{11}+f_{22}+f_{33}$, ${\overline{\Gamma}}=f_{\overline{0}\overline{0}}+f_{\overline{1}\overline{1}}+
f_{\overline{2}\overline{2}}+f_{\overline{3}\overline{3}}$
($f_\mu\equiv\partial f/\partial v_\mu$, $f_{\bar\mu}\equiv\partial f/\partial {\bar v}_{\mu}$ for $\mu=0,1,2,3$), ${\cal L}$
is explicitly given by

\begin{eqnarray}
{\cal L}&=&-\Gamma(\dot{\upsilon}^{2}_{0}+\sum \dot{\upsilon}^{2}_{i})+\overline{\Gamma}(\dot{\overline{\upsilon}}^{2}_{0}+\sum \dot{\overline{\upsilon}}^{2}_{i})+\\ \nonumber
&&\Gamma(\lambda_{0}\dot{\lambda}_{0}+\lambda_{i}\dot{\lambda}_{i})-
\overline{\Gamma}(\overline{\lambda}_{0}\dot{\overline{\lambda}}_{0}+\overline{\lambda}_{i}\dot{\overline{\lambda}}_{i})+\\\nonumber
&&(\epsilon_{ijk}(\overline{\Gamma}_{k}\dot{\upsilon}_{j}+\overline{\Gamma}_{\overline{j}}\dot{\overline{\upsilon}}_{k})+
(\overline{\Gamma}_{0}\dot{\upsilon}_{i}+\overline{\Gamma}_{\overline{0}}\dot{\overline{\upsilon}}_{i}-\overline{\Gamma}_{i}\dot{\upsilon}_{0}-
\overline{\Gamma}_{\overline{i}}\dot{\overline{\upsilon}}_{0}))\overline{\lambda}_{0}\overline{\lambda}_{i}+\\\nonumber
&&(\epsilon_{ijk}(\Gamma_{k}\dot{\upsilon}_{j}+\Gamma_{\overline{j}}\dot{\overline{\upsilon}}_{k})+
(\Gamma_{0}\dot{\upsilon}_{i}+\Gamma_{\overline{0}}\dot{\overline{\upsilon}}_{i}-\Gamma_{i}\dot{\upsilon}_{0}-
\Gamma_{\overline{i}}\dot{\overline{\upsilon}}_{0}))\lambda_{i}\lambda_{0}+\\\nonumber
&&(\epsilon_{ijk}(\Gamma_{\overline{k}}\dot{\upsilon}_{j}+\Gamma_{j}\dot{\overline{\upsilon}}_{k})-
(\overline{\Gamma}_{0}\dot{\overline{\upsilon}}_{i}+\Gamma_{\overline{0}}\dot{\upsilon}_{i}+\overline{\Gamma}_{i}
\dot{\overline{\upsilon}}_{0}+\Gamma_{\overline{i}}\dot{\upsilon}_{0}))\overline{\lambda}_{i}\lambda_{0}+\\\nonumber
&&(\epsilon_{ijk}(\overline{\Gamma}_{k}\dot{\overline{\upsilon}}_{j}+\overline{\Gamma}_{\overline{j}}\dot{\upsilon}_{k})-
(\overline{\Gamma}_{0}\dot{\overline{\upsilon}}_{i}+\Gamma_{0}\dot{\upsilon}_{i}+\overline{\Gamma}_{i}\dot{\overline{\upsilon}}_{0}+
\Gamma_{i}\dot{\upsilon}_{0}))\overline{\lambda}_{0}\lambda_{i}+\\\nonumber
&&\frac{1}{2}(\epsilon_{ijk}(\overline{\Gamma}_{i}\dot{\upsilon}_{0}-\overline{\Gamma}_{0}\dot{\upsilon}_{i}-
\overline{\Gamma}_{\overline{i}}\dot{\overline{\upsilon}}_{0}-\overline{\Gamma}_{\overline{0}}\dot{\overline{\upsilon}}_{i})+
(\overline{\Gamma}_{j}\dot{\upsilon}_{k}-\overline{\Gamma}_{k}\dot{\upsilon}_{j}+\overline{\Gamma}_{\overline{j}}
\dot{\overline{\upsilon}}_{k}-\overline{\Gamma}_{\overline{k}}\dot{\overline{\upsilon}}_{j}))
\overline{\lambda}_{j}\overline{\lambda}_{k}+\\\nonumber
&&\frac{1}{2}(\epsilon_{ijk}(\Gamma_{i}\dot{\upsilon}_{0}-\Gamma_{\overline{0}}\dot{\overline{\upsilon}}_{i}-
\overline{\Gamma}_{0}\dot{\upsilon}_{i}-\Gamma_{\overline{i}}\dot{\overline{\upsilon}}_{0})+(
\Gamma_{k}\dot{\upsilon}_{j}-\Gamma_{j}\dot{\upsilon}_{k}-\Gamma_{\overline{k}}\dot{\overline{\upsilon}}_{j}-
\Gamma_{\overline{j}}\dot{\overline{\upsilon}}_{k}))\lambda_{j}\lambda_{k}-\\\nonumber
&&(\Gamma_{i}\dot{\upsilon}_{j}+\overline{\Gamma}_{i}\dot{\overline{\upsilon}}_{j})(\lambda_{i}\overline{\lambda}_{j}-
\overline{\lambda}_{i}\lambda_{j})+\\\nonumber
&&(\Gamma_{\overline{0}}\dot{\upsilon}_{0}+\overline{\Gamma}_{0}\dot{\overline{\upsilon}}_{0}+\Gamma_{\overline{i}}
\dot{\upsilon}_{i}+\overline{\Gamma}_{j}\dot{\overline{\upsilon}}_{j})(\overline{\lambda}_{0}\lambda_{0}+
\overline{\lambda}_{k}\lambda_{k})+\\\nonumber
&&\epsilon_{ijk}(\Gamma_{\overline{k}}\dot{\upsilon}_{0}+\overline{\Gamma}_{0}\dot{\upsilon}_{k}-
\Gamma_{\overline{0}}\dot{\upsilon}_{k}-\overline{\Gamma}_{k}\dot{\overline{\upsilon}}_{0})\overline{\lambda}_{j}\lambda_{i}+\\\nonumber
&&(\Gamma_{\overline{i}\overline{j}}-\overline{\Gamma}_{ij})\lambda_{0}\overline{\lambda}_{0}\lambda_{i}\overline{\lambda}_{j}+
\epsilon_{ijk}\lambda_{0}\overline{\lambda}_{0}(\overline{\Gamma}_{0k}\lambda_{i}\overline{\lambda}_{j}+
\Gamma_{\overline{0}\overline{k}}\lambda_{i}\overline{\lambda}_{j})+\\\nonumber
&&\lambda_{0}\overline{\lambda}_{0}((\Gamma_{\overline{i}j}\lambda_{i}\lambda_{j}+
\overline{\Gamma}_{i\overline{j}}\overline{\lambda}_{i}\overline{\lambda}_{j})+
\frac{1}{2}\epsilon_{ijk}(\overline{\Gamma}_{0\overline{j}}\overline{\lambda}_{i}\overline{\lambda}_{k}+
\overline{\Gamma}_{j\overline{0}}\overline{\lambda}_{i}\overline{\lambda}_{k}-
\Gamma_{0\overline{k}}\lambda_{i}\lambda_{j}-\Gamma_{i\overline{0}}\lambda_{j}\lambda_{k}))+\\\nonumber
&&\frac{1}{2}\epsilon_{ijk}((\Gamma_{0\overline{0}}+\Gamma_{p\overline{p}})\lambda_{0}\lambda_{i}\lambda_{j}\overline{\lambda}_{k}+
(\overline{\Gamma}_{0\overline{0}}+\overline{\Gamma}_{p\overline{p}})\overline{\lambda}_{0}\overline{\lambda}_{i}\overline{\lambda}_{j}\lambda_{k})+
\\\nonumber
&&(\Gamma_{j\overline{0}}-\Gamma_{0\overline{j}})\lambda_{0}\lambda_{j}\lambda_{i}\overline{\lambda}_{i}-
\frac{1}{2}\epsilon_{ijk}\delta_{pq}(\Gamma_{p\overline{k}}+\Gamma_{k\overline{p}})\lambda_{0}\lambda_{i}\lambda_{j}\overline{\lambda}_{q}+
\\\nonumber
&&(\Gamma_{\overline{j}\overline{0}}-\overline{\Gamma}_{0j})\lambda_{0}\overline{\lambda}_{j}\lambda_{i}\overline{\lambda}_{i}+
\frac{1}{2}\epsilon_{ijk}\delta_{pq}(\overline{\Gamma}_{kp}+
\Gamma_{\overline{k}\overline{p}})\lambda_{0}\overline{\lambda}_{j}\overline{\lambda}_{i}\lambda_{q}+\\\nonumber
&&(\overline{\Gamma}_{0j}-\Gamma_{\overline{0}\overline{j}})\overline{\lambda}_{0}\lambda_{j}\lambda_{i}
\overline{\lambda}_{i}-\frac{1}{2}\epsilon_{ijk}\delta_{pq}(\overline{\Gamma}_{kp}+\Gamma_{\overline{k}\overline{p}})
\overline{\lambda}_{0}\lambda_{j}\lambda_{i}\overline{\lambda}_{q}+\\\nonumber
&&(\overline{\Gamma}_{0\overline{j}}-\overline{\Gamma}_{j\overline{0}})\overline{\lambda}_{0}\overline{\lambda}_{j}\lambda_{i}\overline{\lambda}_{i}-
\frac{1}{2}\epsilon_{ijk}\delta_{pq}(\overline{\Gamma}_{p\overline{k}}+\overline{\Gamma}_{k\overline{p}})
\overline{\lambda}_{0}\overline{\lambda}_{j}\overline{\lambda}_{i}\lambda_{q}+\\\nonumber
&&(\Gamma_{\overline{j}\overline{k}}-\overline{\Gamma}_{jk})\lambda_{k}\overline{\lambda}_{j}\lambda_{i}\overline{\lambda}_{i}+
\frac{1}{2}\epsilon_{ijk}\delta_{pq}(\overline{\Gamma}_{0p}+
\Gamma_{\overline{0}\overline{p}}\lambda_{i}\lambda_{k}\overline{\lambda}_{j}\overline{\lambda}_{q})+\\\nonumber
&&\frac{1}{2}((\Gamma_{k\overline{j}}-\Gamma_{j\overline{k}})\lambda_{j}\lambda_{k}\lambda_{i}\overline{\lambda}_{i}+
(\overline{\Gamma}_{k\overline{j}}-\overline{\Gamma}_{j\overline{k}})\overline{\lambda}_{j}\overline{\lambda}_{k}\overline{\lambda}_{i}\lambda_{i})+
\\\nonumber
&&\frac{1}{4}\epsilon_{ijk}\delta_{pq}((\Gamma_{0\overline{p}}+\Gamma_{p\overline{0}})\lambda_{i}\lambda_{j}\lambda_{k}\overline{\lambda}_{q}-
(\overline{\Gamma}_{p\overline{0}}+\overline{\Gamma}_{0\overline{p}})\overline{\lambda}_{i}\overline{\lambda}_{j}\overline{\lambda}_{k}\lambda_{q})+
\\\nonumber
&&\frac{1}{6}\epsilon_{ijk}((\Gamma_{0\overline{0}}-\Gamma_{p\overline{p}})\lambda_{i}\lambda_{j}\lambda_{k}\overline{\lambda}_{0}-
(\overline{\Gamma}_{0\overline{0}}-\overline{\Gamma}_{p\overline{p}})\overline{\lambda}_{i}\overline{\lambda}_{j}\overline{\lambda}_{k}\lambda_{0})+
\\\nonumber
&&(\Gamma_{\overline{0}\overline{0}}-\overline{\Gamma}_{00})\lambda_{0}\overline{\lambda}_{0}\lambda_{i}\overline{\lambda}_{i}+
\frac{1}{2}\epsilon_{ijk}(\Gamma_{\overline{p}\overline{p}}-\overline{\Gamma}_{00})(\lambda_{0}\lambda_{i}
\overline{\lambda}_{j}\overline{\lambda}_{k}+
\overline{\lambda}_{0}\lambda_{i}\overline{\lambda}_{k}\overline{\lambda}_{j})+\\\nonumber
&&\frac{1}{6}\epsilon_{ijk}((\Gamma_{00}+\Gamma_{pp})\lambda_{i}\lambda_{j}\lambda_{k}\lambda_{0}-
(\overline{\Gamma}_{\overline{0}\overline{0}}+
\overline{\Gamma}_{\overline{p}\overline{p}})\overline{\lambda}_{i}\overline{\lambda}_{j}\overline{\lambda}_{k}\lambda_{0}).
\end{eqnarray}
\par~\par

{\bf {\large Acknowledgments.}}

\par~\par
We are grateful to Sergey Krivonos for useful discussions and to Evgeny Ivanov for the interest in the work and for
pointing out some references to his papers.
A.~N. acknowledges CBPF for hospitality and financial
support during his staying in Rio de Janeiro, where the main part of the present work was carried out.
 He also acknowledges hospitality at the International Center of Theoretical Physics in Trieste, where some part of this work has been done.
 M. G. acknowledges a CLAF grant. Z.~K. acknowledges a FAPERJ grant.
The work was partially supported by the NFSAT-CRDF UC 06/07 (A.~N.,
V.~Y.),
 ANSEF PS1730 (A.~N., V.~Y.), INTAS-05-7928 (A.~N.) grants and by
  Edital Universal CNPq, Proc. 472903/2008-0 (M.G, Z.K., F.T).


\begin{thebibliography}{99}

\bibitem{Ivanov:2003nk}
  E.~Ivanov and O.~Lechtenfeld,
  JHEP {\bf 0309} (2003) 073.


\bibitem{IKL} E.~Ivanov, S.~Krivonos and O.~Lechtenfeld,
  Class.\ Quant.\ Grav.\  {\bf 21} (2004) 1031.
\bibitem{k1}  S.~Bellucci, E.~Ivanov, S.~Krivonos and O.~Lechtenfeld,
  Nucl.\ Phys.\  B {\bf 699} (2004) 226;

 S.~Bellucci, A.~Beylin, S.~Krivonos and A.~Shcherbakov,
  Phys.\ Lett.\  B {\bf 633} (2006) 382;

 S.~Bellucci, S.~Krivonos and A.~Marrani,
  Phys.\ Rev.\  D {\bf 74} (2006) 045005;

 E.~Ivanov,
  Phys.\ Lett.\  B {\bf 639} (2006) 579
  [arXiv:hep-th/0605194].

\bibitem{bbkno}S.~Bellucci, A.~Beylin, S.~Krivonos, A.~Nersessian and E.~Orazi,
  Phys.\ Lett.\   {\bf B616} (2005) 228.
\bibitem{kno}S.~Krivonos, A.~Nersessian and  V.~Ohanyan
 {\em Phys.\ Rev.\/} {\bf D75} (2007) 085002.


\bibitem{bko} S.~Bellucci, S.~Krivonos and V.~Ohanyan,
  Phys.\ Rev.\   {\bf D76} (2007) 105023.
\bibitem{freedom} S.~Bellucci and A.~Nersessian,
  Phys.\ Rev.\  D {\bf 73} (2006) 107701.

\bibitem{Z} D. Zwanziger, {\em Phys. Rev.\/} {\bf 176}
(1968) 1480.

\bibitem{mic}H. McIntosh and A. Cisneros, {\em J. Math. Phys.\/} {\bf 11}
(1970) 896.
%
%
%
%
\bibitem{Gates}
  S.~J.~J.~Gates and L.~Rana,
  Phys.\ Lett.\  B {\bf 342} (1995) 132.

\bibitem{pashtop}A.~Pashnev and F.~Toppan,
  J.\ Math.\ Phys.\  {\bf 42} (2001) 5257.


\bibitem{root} S.~Bellucci, S.~Krivonos, A.~Marrani and E.~Orazi,
  Phys.\ Rev.\  D {\bf 73} (2006) 025011.

  \bibitem{ks}
P.~Kustaanheimo,~E.~Stiefel,~J.~Reine Angew Math.,
 {\bf 218} (1965) 204.


\bibitem{bny} S.~Bellucci, A.~Nersessian, A.~Yeranyan,  Phys.\ Rev.\ {\bf D 74} (2006) 065022

\bibitem{ntt}A.~Nersessian, V.~Ter-Antonian,  M.~M.~Tsulaia, Mod.\ Phys.\ Lett.\  {\bf A11} (1996), 1605.
\bibitem{iwai}T.~Iwai,
 J.\ Geom.\ Phys. {\bf 7} (1990) 507;
L.~G.~Mardoyan,
 A.~N.~Sissakian, V.~M.~Ter-Antonyan, Phys.\ Atom.\ Nucl.
 {\bf 61} (1998) 1746.

\bibitem{4h}S.~C.~Zhang, J.~P.~Hu,
Science {\bf 294} (2001) 823.

\bibitem{hasebe}  K.~Hasebe,
  Phys.\ Rev.\ Lett.\  {\bf 94} (2005) 206802.

\bibitem{ped}  C.~Pedder, J.~Sonner and D.~Tong,
  JHEP {\bf 0803}, 065 (2008)


\bibitem{lnp} A.~Nersessian,
  Lect.\ Notes Phys.\  {\bf 698} (2006) 139.


\bibitem{mny}
  L.~Mardoyan, A.~Nersessian and A.~Yeranyan,
  Phys.\ Lett.\  A {\bf 366} (2007) 30.

\bibitem{schw} J.~S.~Schwinger,
  Science {\bf 165} (1969) 757.

\bibitem{gauging}
  F.~Delduc and E.~Ivanov,
  Nucl.\ Phys.\   {\bf B753} (2006) 211;
  {\it ibid.} {\bf B770} (2007) 179;
  {\it ibid.} {\bf B787} (2007) 176.

\bibitem{horv}
  C.~Duval and P.~Horvathy,
  Annals Phys.\  {\bf 142} (1982) 10;

   P.~A.~Horvathy and J.~P.~Ngome,
  arXiv:0902.0273 [hep-th].


\bibitem{oku} S. Okubo, J.\ Math.\ Phys.\ {\bf 32} (1991) 1657.

\bibitem{kt} Z. Kuznetsova and F. Toppan, JHEP05 (2005) 060.

\bibitem{krt} Z. Kuznetsova, M. Rojas and F. Toppan, JHEP03 (2006) 098.

\bibitem{kt2} L. Carvalho, Z. Kuznetsova and F. Toppan, {\it Supersymmetric extensions of the Hopf maps}, {work in preparation}.

\bibitem{sm1} A.V. Smilga, Sov. Phys. JETP {\bf 64} (1986) 8.

\bibitem{CR} M. de Crombrugghe and V. Rittenberg, Ann. Phys. {\bf 151} (1983) 99.

\bibitem{sm2} A.V. Smilga, Nucl. Phys. {\bf B 291} (1987) 241.

\bibitem{grt} M. Gonzales, M. Rojas and F. Toppan, preprint CBPF-NF-026/08, arXiv:0812.3042[hep-th].
To appear in Int. J. Mod. Phys. A.
%
\bibitem{Ivanov:2007sh}
  E.~Ivanov, O.~Lechtenfeld and A.~Sutulin,
  Nucl.\ Phys.\  B {\bf 790}, (2008) 493.

\end{thebibliography}
\end{document}